> "Existing theories about the behavior of stock prices are remarkably inadequate."
> G. Soros[1].

# Non - Randomness Stock Market Price Model.
## Aleksey Kharevsky.

Phenomenology. Even by the first look at the tick by tick price chart one can see that a set of price values makes a space of isolated points: for every point there is a region that does not contain any other points. This phenomenon does not allow using the continuity concept.

There are many reasons why it is also wrong to use the concept of randomness. The first one is that we do not know what randomness is. The test with a coin toss is a classic in the discussion of randomness. It is generally regarded that heads and tails are the equally possible outcomes, and edge is almost impossible. But where does randomness disappear if a coin falls on the surface densely packed with billiard balls? The relative properties of the space in which the coin will get "stuck" is a major factor. The above said is fully applicable to price movements. The space where price function takes its values is non-uniform and non-isotropic.

While price behavior is seemingly chaotic and unpredictable its change understood as the variation of the function is finite for any time interval.

1. Formulation of the problem. Based on the above observations it is suggested to look at price as an everywhere discontinuous[2] time function of bounded variation; denote it as $f(t)$, $t \in T = [a,b] \subset \mathbb{R}$. Let's denote the difference between the function values at the endpoints of $T$ as $D = f(b) - f(a)$. We are interested in the conditions and the number of functions under which $D \leq 0$. The main tool used to construct the model is the obvious property of Jordan decomposition.

---

[1] G. Soros. The Alchemy of Finance. John Wiley & Sons, Inc. NY, 1987.
[2] Р. Бэр. Теория разрывных функций. Перевод с фр. и ред. А. Я. Хинчина. ГТТИ, М.-Л., 1932.

2. Hyperbolic property of Jordan decomposition. We have the following lemma: for an arbitrary function of bounded variation the following equalities hold true

$$\Sigma^+ - \Sigma^- = D,$$
$$\Sigma^+ + \Sigma^- = V.$$
(1.1)

Here $V = \sup \sum_{k=1}^{n} |f(t_k) - f(t_{k-1})|$ is total variation of the function $f(t)$ over segment $T$, $\Sigma^+ = f^+(b) - f^+(a)$, $\Sigma^- = f^-(b) - f^-(a)$. Functions $f^+(t)$ and $f^-(t)$ are monotonic non-decreasing functions in Jordan decomposition $f(t) = f^+(t) - f^-(t)$. The proof is trivial if the following representations are used for these functions $f^+(t) = [V(t) + f(t)]/2$, $f^-(t) = [V(t) - f(t)]/2$. Multiplication of the equalities of the lemma defines a hyperbola, hence the name of the property.

3. The variation of a discontinuous function. If a function has a discontinuity at a point, the oscillation of the function at this point is positive; it can be represented as $\omega_k = M_k - m_k$, where $\omega_k$, $M_k$, $m_k$ are oscillation, least upper bound and greatest lower bound of the function at the point of discontinuity $t_k \in T$, respectively. For the terms of the total variation we have

$$|f(t_k) - f(t_{k-1})| = \max(M_{k-1}, M_k) - \min(m_{k-1}, m_k) =$$

$$= \frac{M_{k-1} + M_k + |M_{k-1} - M_k|}{2} - \frac{m_{k-1} + m_k - |m_{k-1} - m_k|}{2} =$$

$$= \frac{\omega_{k-1} + \omega_k + |M_{k-1} - M_k| + |m_{k-1} - m_k|}{2}.$$

Further, on one hand we have

$$\frac{\omega_{k-1} + \omega_k + |M_{k-1} - M_k| + |m_{k-1} - m_k|}{2} =$$

$$= \frac{\omega_{k-1} + \omega_k + |M_{k-1} - m_{k-1} + m_{k-1} - M_k| + |M_k - M_k + m_{k-1} - m_k|}{2} =$$

$$= \frac{\omega_{k-1} + \omega_k + |\omega_{k-1} + m_{k-1} - M_k| + |\omega_k - M_k + m_{k-1}|}{2} \le \omega_{k-1} + \omega_k + |m_{k-1} - M_k|.$$

On the other hand

$$\frac{\omega_{k-1}+\omega_k+|M_{k-1}-M_k|+|m_{k-1}-m_k|}{2}=$$

$$=\frac{\omega_{k-1}+\omega_k+|M_{k-1}-m_k+m_k-M_k|+|M_{k-1}-M_{k-1}+m_{k-1}-m_k|}{2}=$$

$$=\frac{\omega_{k-1}+\omega_k+|M_{k-1}-m_k-\omega_k|+|M_{k-1}-m_k-\omega_{k-1}|}{2}\le \omega_{k-1}+\omega_k+|m_k-M_{k-1}|.$$

From the above we obtain total variation

$$V=\sum_{k=1}^{n}[\omega_{k-1}+\omega_k]+\sum_{k=1}^{n}\left[\min\left(|M_k-m_{k-1}|,|M_{k-1}-m_k|\right)\right]. \qquad (1.2)$$

Consider the first term. The set of discontinuities of the function $f(t)$ is at most countable. It follows from the definition of oscillation that for any point of discontinuity $t_k$ and any positive number $\varepsilon>0$ there is an elementary segment $\Delta_k, t_k \in \Delta_k$ for which $0 \le \omega(\Delta_k)-\omega_k < \varepsilon$. This way the set of elementary segments is at most countable. For the variation to be bounded there needs to exist a finite set of elementary segments that would, in a different from the classical understanding sense, cover segment $T$. Let's represent the oscillation over the elementary segment in the form of

$$\omega(\Delta_k)=\lambda\cdot\rho_k,$$

where $\rho_k=\sum_{j=1}^{\infty}\dfrac{\rho_{j,k}}{2^j}$, $k\in\mathbb{N}$ and $\rho_{j,k}$ can take value from the set $\{0,1\}$. Number $\rho_k$ can be called density of oscillation at the elementary segment. We can state that the set of discontinuities at the elementary segment is at most countable and oscillation is bounded by $\lambda$. Over segment $T$ the following condition holds true

$$\frac{|\omega(\Delta_k)-\omega(\Delta_{k-1})|}{\lambda}=|\rho_k-\rho_{k-1}|<\varepsilon_\rho, \qquad (1.3)$$

where $0<\varepsilon_\rho<1$. Almost all sums $\rho_k$ fulfill this condition. For these sums with precision of up to number $\varepsilon_\rho\lambda$ the following equality is true $\omega(\Delta_k)=\omega(\Delta_{k-1})$, hence $\Delta_k=\Delta_{k-1}=\Delta$ and the total variation is bounded because $n=(b-a)/\Delta$ is finite.

Consider the second term in (1.2). This term complements the total variation of function with the help of the rules of transition between the neighboring oscillations; the rules, in turn, are dictated by the definition for the least upper bounds of the number set. The rules are explained by the diagrams $A$ and $B$.

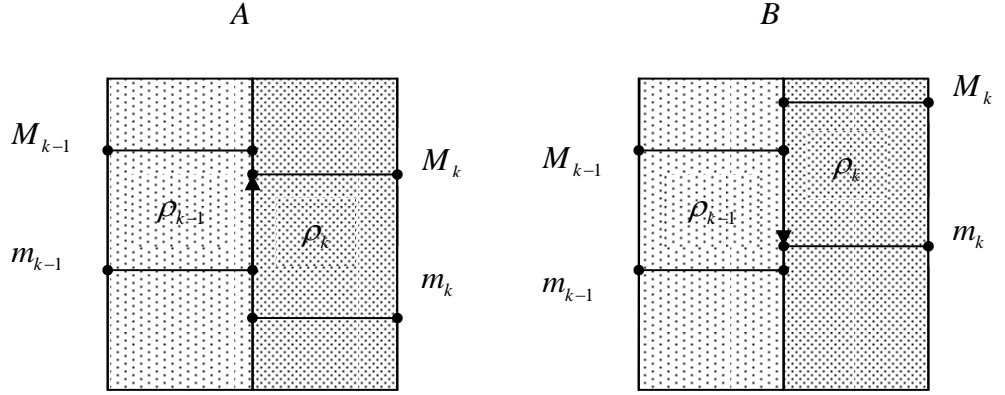

With that said, formula (1.2) can be written as follows

$$V = 2n\lambda\rho(1+\alpha_1+\alpha_2).  \qquad (1.4)$$

Here the parameter $\rho = \dfrac{1}{n}\cdot\sum_{k=1}^{n}\sum_{j=1}^{\infty}\dfrac{\rho_{j,k}}{2^j}$ is the average density of oscillations over $T$. The parameter $|\alpha_1| = \left|\dfrac{\rho_0 - \rho_n}{2n\rho}\right| < \dfrac{\varepsilon_\rho}{2\rho}$ is the relative deviation of density of oscillation at the ends of $T$. The parameter $\alpha_2 = \dfrac{1}{2n\lambda\rho}\sum_{k=1}^{n}\left[\min\left(|M_k - m_{k-1}|, |M_{k-1} - m_k|\right)\right]$ is the average shift of oscillations relative to each other over $T$. We can say that the parameters characterize non-uniformity (a different density) and anisotropy (prescribed direction for the transitions) of the target space of the function $f(t)$. We combine them into one parameter $\alpha = \alpha_1 + \alpha_2$. In the common case $-\infty < \alpha < \infty$.

4. The case of $\alpha = 0$. Bringing the relation (1.1) to the average oscillation $\omega = \lambda\rho$, we get

$$\Sigma_\omega^+ - \Sigma_\omega^- = d,$$
$$\Sigma_\omega^+ + \Sigma_\omega^- = 2n.$$

Here $\Sigma_\omega^+ = \Sigma^+/\omega$, $\Sigma_\omega^- = \Sigma^-/\omega$, $d = D/\omega$. Let's represent numbers $\Sigma_\omega^+$ and $\Sigma_\omega^-$ as whole and fractional parts $\Sigma_\omega^+ = Q^+ + q^+$, $\Sigma_\omega^- = Q^- + q^-$. Their sum is a natural number, so $q^+ + q^- = 1$, and it is also an even number, so $Q^+ - Q^- = 2z + 1$, where $z \in \mathbb{Z}$. Then $\Sigma_\omega^+ - \Sigma_\omega^- = 2z + 2q^+ = d$. For the value of $D$ to be zero $q^+$ must be zero. Then we have $d = 2z$. Ultimately we get the main equalities

$$\begin{aligned}\Sigma_\omega^+ - \Sigma_\omega^- &= 2z, \\ \Sigma_\omega^+ + \Sigma_\omega^- &= 2n.\end{aligned} \qquad (1.5)$$

Here $n \in \mathbb{N}$, $z \in \mathbb{Z}$, $|z| \leq n$. The relations (1.5) determine the number of functions over the segment $2n$ that have the difference $z$ (the difference is expressed in the average oscillation $\omega$). It follows from the equalities (1.5) that the number of functions having $\Sigma_\omega^+$ differences $z$ is the number of $\Sigma_\omega^+$ combinations from the set of $2n$ elements. Adding up the equalities we get $\Sigma_\omega^+ = z + n$. The total number of the functions that satisfy the equalities equals $2^{2n}$. Therefore the density of distribution of the functions that have the difference $z$ equals

$$p(z) = \frac{C_{2n}^{z+n}}{2^{2n}} \approx \frac{e^{-\frac{z^2}{n}}}{\sqrt{\pi n}} = \frac{1}{\sqrt{2\pi}} e^{-\frac{\zeta_z^2}{2}} \cdot \Delta\zeta.$$

Here the following notation is used $\zeta_z^2 = 2z^2/n$, $\Delta\zeta = (\zeta_{z+1} - \zeta_z)$. For large values of number $n$ the calculation of the relative number of functions with the difference not more than $\zeta$, denote it as $P(\zeta)$, leads to the Gauss integral:

$$P(\zeta) = \Phi(\zeta) = \frac{1}{\sqrt{2\pi}} \int_{-\infty}^{\zeta} e^{-\frac{x^2}{2}} dx.$$

5. The case of $\alpha \neq 0$. The segment $[-n, n]$ contains a set of values $z$ in the case of $\alpha = 0$. To save the number of oscillations for the general case of $\alpha \neq 0$ we have to transfer it into segment $[-n + z_0, n + z_0]$, $z_0 \in \mathbb{Z}$. The range of values for the differences $z$ will be determined in this case by the set $[-n, n] \cap [-n + z_0, n + z_0]$ which is equivalent to the inequality

$$|2z - z_0| \leq 2n - |z_0|. \qquad (1.6)$$

This inequality gives a hint for the substitution of variable; we put

$$\begin{aligned}2z' &= 2z - z_0, \\ 2n' &= 2n - |z_0|.\end{aligned}$$

Under the new variables we have a system of equalities

$$\Sigma_\omega^+ - \Sigma_\omega^- = 2z',$$
$$\Sigma_\omega^+ + \Sigma_\omega^- = 2n'.$$

Here $n' \in \mathbb{N}$, $z' \in \mathbb{Z}$, $|z'| \leq n'$. Now it is possible to apply the result obtained earlier by analogy. The relative number of the functions with difference of not more than $\zeta'$ equals to $\Phi(\zeta')$. From the variable substitution formulas it follows that $z_0 = -2n\alpha$ as well as the formula of transition from variable $\zeta'$ to variable $\zeta$

$$\zeta' = \frac{\zeta + \alpha\sqrt{2n}}{\sqrt{1-|\alpha|}}. \tag{1.7}$$

We are ready to find the relative number of the functions under which of $D \leq 0$ or $\zeta \leq 0$. Let's denote this number as $P(\zeta \leq 0)$. Now we have the solution

$$P(\zeta \leq 0) = \Phi\left(\frac{\alpha\sqrt{2n}}{\sqrt{1-|\alpha|}}\right), \tag{1.8}$$

where $-1 < \alpha < 1$ and $n \in \mathbb{N}$.

7. The Discussion. There are two special cases. In the case of $\alpha = 0$ we have the value of $P(\zeta \leq 0)$ equal to $1/2$. In the case of $|\alpha| \to 1$ the value of $P(\zeta \leq 0)$ may be equal to $0$ or to $1$ depending on the sign of $\alpha$. In fact the value of $\alpha$ depends of the frame of reference in which the observer is located. We will call the frame of reference with $\zeta'$ variable the absolute frame of reference (AFR) and with $\zeta$ variable the relative frame of reference (RFR).

One may have noticed that the Model is very similar to probability models. Perhaps parameter $\alpha$ is responsible for the distribution law. A good test for the Model is to obtain the distribution law of the heavy tails[3] type. The tail of the distribution means $\Phi(\zeta)$ if $\zeta < 0$ and $1-\Phi(\zeta)$ if $\zeta > 0$. Then we obtain $\Phi(\zeta') - \Phi(\zeta) > 0$ for any $\zeta < 0$ and $\alpha > 0$; $\Phi(\zeta) - \Phi(\zeta') > 0$ for any $\zeta > 0$ and $\alpha < 0$. Assume that $\alpha\sqrt{2n} \ll 1$ and that it is a nonlinear function of $\zeta$. With the help of the Taylor approximation of order two and keeping in mind that for heavy tails $sign(\alpha) = -sign(\zeta)$ one can find that

$$\alpha\sqrt{2n} = -\frac{\zeta}{|\zeta|} \cdot \left(C_1 + C_2 \frac{\zeta^2}{2}\right).$$

---

[3] B. Mandelbrot. The variation of certain speculative prices. The Journal of Business, Vol. 36, No. 4 (Oct., 1963), pp. 394-419.

We find the constants $C_1$ and $C_2$ by using the fact that AFR is invariant to $\alpha$. From the AFR viewpoint $\zeta'$ obeys the rule of "three sigma" at the same time from the RFR viewpoint heavy tails are observed up to six sigma. Anomalies are also observed near zero values. When $\zeta' = 0$ in AFR an observer in RFR can see a small $\zeta_0$. Using formula (1.7) and the approximation we have the equations

$$C_1 + 18C_2 = 3,$$
$$C_1 + \frac{\zeta_0^2}{2}C_2 = \zeta_0.$$

For heavy tails distribution we have

$$P(\zeta)\big|_{\alpha\sqrt{2n}\ll 1} \approx \Phi\left[\zeta - \frac{\zeta}{12|\zeta|}(\zeta^2 - \zeta_0^2)\right].$$

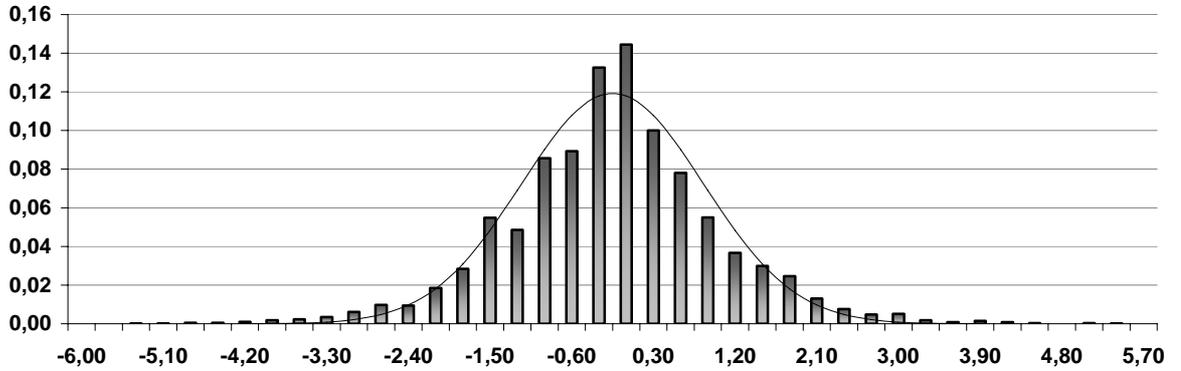

The figure shows a typical fat tails distribution. For modeling the value of $\zeta_0$ a pseudo random number from 0 up to 1 was used. The solid line corresponds to the normal law and the grey bars correspond to the fat tails model.

The analogy can be extended. Using the formula (1.7) and the definition for the function $\Phi(\zeta')$, we can find the statistical content of the parameter $\alpha$

$$\begin{aligned}\mu_\omega &= -2n\alpha, \\ \sigma_\omega^2 &= 2n(1-|\alpha|).\end{aligned} \qquad (1.9)$$

Here the following notation is used $\mu_\omega = \mu/\omega$, $\sigma_\omega = \sigma/\omega$. Numbers $\mu$ and $\sigma$ are the mean and the standard deviation of $D$. From the formulas (1.9) it immediately follows that

$$P(\zeta \leq 0) = \Phi(-\mu/\sigma). \qquad (1.10)$$

The formula (1.10) may be used in technical analysis. By the way the formula (1.6) is a prototype for Bollinger bands. Finally here is a comparative table.

| Theory of probability | This Model. |
|---|---|
| Random event. | Elementary segment $\Delta_k$. |
| Random value. | Oscillation $\omega(\Delta_k)$. |
| Random process. | Function fulfilled to (1.5). |
| Independency. | The condition (1.3). |

The full analysis is not completed yet. The new results will be published as they come available.

References.
1. G. Soros. The Alchemy of Finance. John Wiley & Sons, Inc. NY, 1987.
2. Р. Бэр. Теория разрывных функций. Перевод с фр. и ред. А. Я. Хинчина. ГТТИ, М.-Л., 1932.
3. B. Mandelbrot. The variation of certain speculative prices. The Journal of Business, Vol. 36, No. 4 (Oct., 1963), pp. 394-419.